\documentclass[preprint]{revtex4-1}
\usepackage{graphicx}

\usepackage{natbib}

\newcommand{\be}{\begin{equation}}
\newcommand{\ee}{\end{equation}}
\newcommand{\nn}{\mbox{} \nonumber \\ \mbox{} }
\newcommand{\ba}{\begin{eqnarray}}
\newcommand{\ea}{\end{eqnarray}}

\newcommand{\Alfven}{Alfv\'{e}n\,}

\newcommand\etal{\textit{et al.}}
\newcommand\eg{\textit{e.g.}}
\newcommand\cf{\textit{cf.}}

\newcommand{\Bf}{{magnetic field}}
\newcommand{\Bfs}{{magnetic fields}}

\newcommand{\NS}{neutron star}

\newcommand{\mnras}{MNRAS}

\newcommand{\physrep}{Phys. Rep.}

\begin{document}

\title{Simple waves in relativistic fluids}

\author{Maxim Lyutikov\\
Department of Physics, Purdue University, \\
 525 Northwestern Avenue,
West Lafayette, IN
47907-2036 }

\begin{abstract}
 We consider the Riemann problem for  relativistic flows of polytropic fluids  and find relations for the flow characteristics. Evolution of physical quantities take especially simple form for the case of   cold magnetized plasmas.  We    find exact, explicit  analytical  solutions for one dimensional  expansion of magnetized  plasma into vacuum, valid for   arbitrary magnetization.   We also consider  expansion into  cold unmagnetized  external medium both for stationary initial conditions and for initially moving plasma, as well as reflection of rarefaction wave from a wall. 
We also find self-similar structure of three-dimensional  magnetized  outflows into vacuum, valid close to the plasma-vacuum interface.

The key results of this work, the self-similar solutions,  were incorporated  post-initial submission   into   appendices of the published version of  Granot \etal\  (2010).  

 \end{abstract}
\maketitle

\section{Introduction}

Relativistic shock waves are common in different physical systems \citep{2003LRR.....6....7M}, from heavy ion nuclear collision \citep[\eg][]{2009PhRvL.103c2301B} to astrophysical shocks in pulsar winds \citep{kc84},  Active Galactic Nuclei \citep[\eg][]{Krolik:1999} and Gamma Ray Bursts \citep{PiranReview}. 
Many modern computational algorithms are based on the  solution of Riemann problems \citep[\eg][]{Toro}. These algorithms are based on Godunov-type shock-capturing schemes  and do not require large artificial viscosity or smoothing operators.    Analytical solutions to the corresponding Riemann problems are then important for code testing. 

  Exact, explicit  non-linear solutions of relativistic fluid equations, and especially  relativistic MHD equations, are rare.  In a general form the relativistic Riemann problem was solved by \cite{1994JFM...258..317M,Romero05}, who find the solutions for Riemann invariants and for the characteristics in quadratures.
  In this paper we find simple expressions for the characteristics of Riemann simple waves, and in particular for the astrophysical important case of dynamics of a cold, relativistically magnetized plasma.  Our  results can be used for  benchmark estimates of the overall dynamical behavior in   numerical simulations of relativistic flows and  strongly magnetized outflows in particular.

 \section{One dimensional expansion of  polytropic gas  into vacuum}
 \label{Simple}

  \subsection{ Polytropic equations of state}

Let us  assume that  pressure is a polytropic function of density, $P =K  \rho^\Gamma$, with constant $\Gamma$; then the internal energy density, excluding rest mass is ${\cal E}=P/(\Gamma -1)$, enthalpy $h = \rho + {\cal E} +P = \rho+\Gamma/(\Gamma-1)P$ and  sound speed can be expressed as
   \be 
   c_s^2 = {\rho \over h} {\partial P \over \partial {\rho}} = {(\Gamma-1) \Gamma  P/\rho \over (\Gamma-1)  + \Gamma P/\rho}
   \label{cccc}
   \ee
   The corresponding four-velocity is
   \be
   u_s^2 = {c_s^2 \over 1- c_s^2} = {(\Gamma-1) \Gamma  P/\rho  \over (\Gamma-1)  +  \Gamma (2-\Gamma)  P/\rho }
   \label{us}
   \ee
%   Non-relativistic case corresponds to $P \ll \rho$, $u_s^2 =\Gamma P/\rho$.

%Using EoS and definition of sound speed, we can express density as a function of sound speed,
%\be
%\rho = \left( K  \left({\Gamma \over \Gamma-1}\right) {\Gamma-1-c_s^2 \over c_s^2} \right)^{1\over \Gamma-1}
%\label{rho}
%\ee
%The case $c_s^2= \Gamma-1$, allowed for $\Gamma \leq 2$, corresponds to $\rho=0$. Naturally, in this case we should set $\Gamma= 4/3$, EoS of a photon gas.
% An exception is a case of  $\Gamma=2$, which  corresponds to cold strongly magnetized plasma moving in a direction perpendicular to the \Bf. 

 \subsection{Riemann invariants}
 
Consider a one-dimensional  flow of fluid  along $z$ direction, neglecting variations of quantities across the flow.
The governing 
equations are
\ba &&
\partial_t( \gamma \rho) + \partial_z (\gamma \beta \rho) =0
\nn &&
\partial_t T_{00} + \partial_z T_{0z}=0
\nn &&
\partial_tT_{0z}+  \partial_z T_{zz}=0
\nn &&
T_{00}= \gamma^2 ({\cal E}+P + \rho) - P
\nn &&
T_{0z}=  \gamma^2 \beta ({\cal E}+P + \rho)
\nn &&
T_{zz}=  \gamma^2 \beta^2 ({\cal E}+P + \rho)+P
\label{main0}
\ea
where $P$ is pressure, ${\cal E}$ is energy density (excluding rest mass), $\rho$ is density and $\beta$ and $\gamma$ are fluid's velocities and Lorentz factors.

\begin{figure}[h]
 \begin{center}
\includegraphics[width=.99\linewidth]{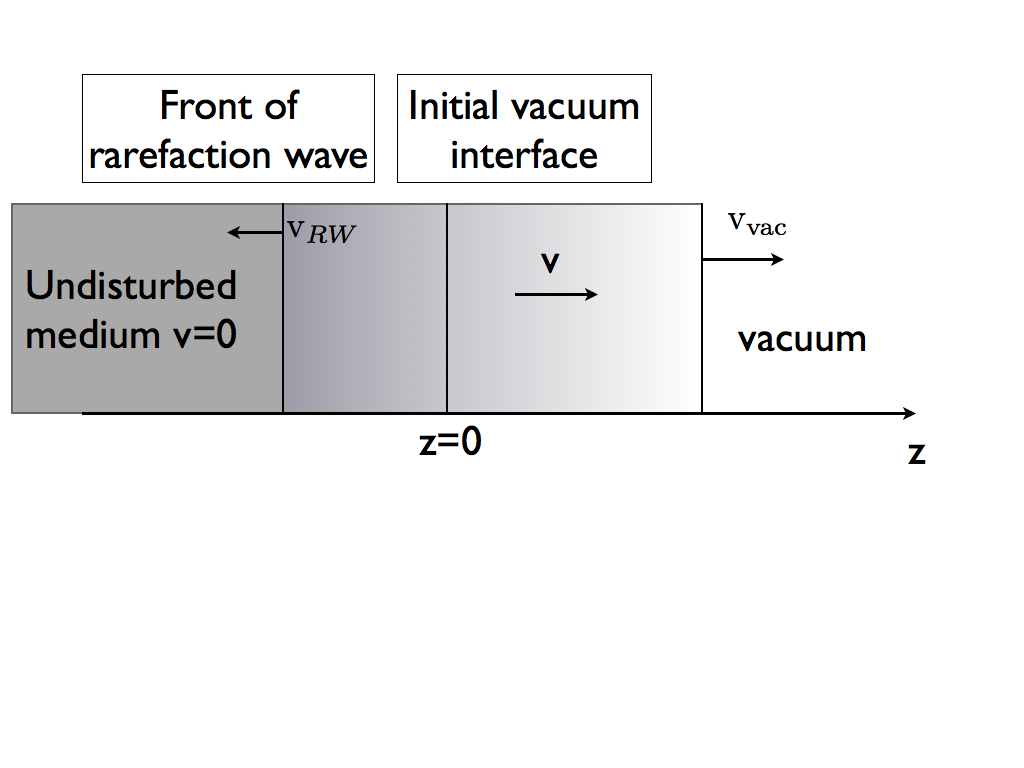}
\end{center}
\vskip -1 truein
\caption{Geometry of the flow. Initially   the fluid  at rest occupies region $z<0$ with constant density and pressure  and  that expansion proceeds into positive $z$ direction. The fluid moves in the positive z-direction while the front of the rarefaction wave propagates in the negative $z$ direction. }
\label{Pcicture-Expan}
\end{figure}

Assume that initially the fluid  occupies region $z<0$ with constant density and pressure  and  that expansion proceeds into positive $z$ direction, Fig \ref{Pcicture-Expan}.

Combining the conservation of energy and momentum,  we find \citep{1948PhRv...74..328T,JohnsonMcKee}
\ba &&
\left (\partial _t + \beta \partial_z \right) \beta= - {  \left (\beta \partial _t + \partial_z \right)P \over({\cal E}   +\rho+P) \gamma^2}
\nn &&
 \left ( \partial _t + \beta \partial_z \right)({\cal E}   +\rho)= - ({\cal E}   +\rho+P) \gamma^2  \left(\beta \partial _t +  \partial_z \right)  \beta
 \label{Eq0}
\ea
These two equations  and  the equation of state combine to give equations for
Riemann invariants
\ba &&
 \left( \partial_t + {\beta+ \beta_{c_s} \over 1+\beta_{c_s} \beta} \partial_z \right) J_+ =0
\nn &&
\left( \partial_t + {\beta- \beta_{c_s} \over 1- \beta_{c_s} \beta} \partial_z\right) J_- =0
\nn && 
J_{\pm} = {1\over 2} \ln { 1+\beta \over 1-\beta}\pm {1 \over \sqrt{\Gamma-1} } \ln { \sqrt{\Gamma-1}   + c_s\over  \sqrt{\Gamma-1}  - c_s}
\label{J}
\ea
\citep{1994JFM...258..317M}. ($J_+ $ corresponds to the forward characteristic). 
The combinations $ {(\beta\pm \beta_{c_s} )/( 1\pm \beta_{c_s} \beta)}$ can be recognized as relativistic transformation of velocities, plasma velocity and wave phase velocity.     Riemann invariants are constant along characteristics
\be
d_t z_\pm= {\beta \pm \beta_{c_s} \over 1\pm \beta \beta_{c_s}}
\ee
(the forward characteristics correspond to upper sign).

Equations (\ref{J}) simplify if we introduce Doppler factor
 \ba &&
 \delta_\beta = {1\over (1-\beta) \gamma }= \sqrt{1+\beta\over 1-\beta}
 \nn &&
 \delta _\eta = \sqrt{ 1 + \eta \over 1-\eta}
 \nn &&
  \delta _{c_s} = \sqrt{ 1 +c_s \over 1- c_s}
 \ea
 (Under co-linear Lorentz transformation, the corresponding Doppler factors change as $\delta  \rightarrow \delta' \delta_{\rm boost}^{\pm 1}$,
where $\delta'$ is a Doppler factor in a frame moving with respect to the observer frame with Doppler factor $\delta_{\rm boost}$. Doppler factors multiply for aligned velocities or divide in case of counter aligned velocities.) 

Next we assume that all quantities depend on self-similar coordinate $\eta=z/t$. 
 When expressed in terms of Doppler factors, Eqns. (\ref{J}) then  simplify
\ba && 
(\delta_{c_s} ^2 \delta_\beta^2 - \delta _\eta^2 ) \partial_\eta J_+=0
\label{8a} \\
&&
( \delta_\beta^2 - \delta _\eta^2  \delta_{c_s} ^2) \partial_\eta J_-=0
\label{8b} 
\ea
%The condition on $J_+$ at the initial moment ($\beta =0$, $\beta_A = \sqrt{\sigma /(1+\sigma)}$  immediately gives the expansion  velocity of the plasma-vacuum interface,
% where $\beta_A=0$, 
 %\be
 % \beta_f ={ 2 \sqrt{ \sigma (1+\sigma)} \over 1+2 \sigma}, \, \gamma_f = 1+ 2 \sigma
 % \ee
 This gives $\delta_\beta= \delta_\eta   \delta_{c_s}$ for the forward   and  $\delta_\beta=  \delta_\eta   / \delta_{c_s}$  for the backward characteristics correspondingly.

Eq. (\ref{8b})   is  identically satisfied on the forward characteristics, while Eq. (\ref{8a}) is satisfied  on the backward characteristics.
Using the expressions  for Riemann invariants and the shape of characteristics we find
\ba &&
\partial_\eta (J_+ ( \delta_\beta =  \delta_{c_s} \delta_\eta))=0
\nn &&
\partial_\eta (J_- ( \delta_\beta =   \delta_\eta/\delta_{c_s}))=0
\ea
Which gives the full solution for simple waves in relativistic fluids:
\ba &&
\delta_{\eta,+} = {C_0 \over \delta _{c_s }} \left( { 1-c_s /\sqrt{\Gamma-1}\over 1+c_s /\sqrt{\Gamma-1} }\right)^{1/\sqrt{\Gamma-1}}, \, \delta_ \beta = \delta_{ c_s} \delta_\eta, \, 
\mbox{forward characteristics}
\\ &&
\delta_{\eta,-} = {C_0 \delta_{c_s} } \left( { 1+ c_s /\sqrt{\Gamma-1}\over 1-c_s /\sqrt{\Gamma-1} }\right)^{1/\sqrt{\Gamma-1}}, \, \delta_ \beta = \delta_\eta/  \delta _{c_s}, \, 
\mbox{backward characteristics}
\label{general}
\ea
where $C_0$ is a constant to be determined from initial condition.
Relations (\ref{general}) give a general relation between the self-similar coordinate $\eta$, local sound speed in the flow $c_s$  and velocity of the flow $\beta$.  
These are transcendental equations for $c_s(\eta)$ and  $\beta(\eta)$

\subsection{Magnetized cold plasma}

In many astrophysical phenomena the magnetic field controls the overall dynamics
of plasma.  These likely include  magnetars (strongly magnetized {\NS}s possessing
super-strong \Bfs), pulsars and pulsar wind nebulae, jets of Active Galactic
Nuclei and Gamma-Ray Bursters. The plasma of these exotic objects can be described as
relativistically strongly magnetized. This means that  the inertia of this
plasma is dominated by the \Bf\ and not by the particle rest mass, $B^2/ 8 \pi
\gg \rho c^2$, and that  the propagation speed of \Alfven waves approaches the speed
of light. Thus, the conditions in such plasmas  are very different  from the conditions encountered in 
laboratory plasmas, plasmas of planetary magnetospheres, and the interplanetary
plasma.

For example, 
magnetic fields  may play an important dynamical role in the Gamma Ray Burst  outflows \cite[\eg][]{LyutikovJPh,Lyutikov:2009}. They may power the relativistic outflow through, \eg, Blandford-Znajek \citep{BlandfordZnajek} process \citep[see also][]{Komissarov05}, and contribute to particle acceleration in the emission regions. The extreme physical conditions of magnetically dominated plasma suggest new
physics and demand a systematic study.  In relativistic magneto-plasma Godunov-type schemes are discussed by  \citep{1999MNRAS.303..343K}.

 Consider  cold ideal magnetized plasma moving perpendicular to the direction of  the \Bf.   This case, in fact, is  just  a particular case of polytropic equations of state $P \propto \rho^\Gamma$  for unmagnetized fluid flow \citep{LLVIII}.  The case of cold magnetized plasma  is, in fact, somewhat special since it gives   simple relations between density and sound four-velocity, as we demonstrate below.
For cold plasma, when the pressure is purely magnetic, $P= B^2/2$ (for convenience, we renormalize \Bf\ by $\sqrt{4 \pi}$ below),  polytropic  index $\Gamma=2$, and defining  magnetization parameter $\sigma= B^2 /\rho$, Eq. (\ref{us}) gives
\be
U_s \equiv U_A= {\beta_A \over \sqrt{1-\beta_A^2}} =   \sqrt{  \sigma}
\ee
the four-velocity of \Alfven and fast magnetosonic waves in strongly magnetized plasma \citep[][$\beta_A$ is \Alfven velocity]{kc84}. 

In case of magnetized plasma, the equations of motion (\ref{main0}) should be supplemented with induction equation and definitions of magnetic energy density and pressure
\ba &&
\partial_t( \gamma B) + \partial_z (\gamma \beta B) =0
\nn &&
{\cal E}=P = B^2/2
\label{main00}
\ea
where  $B$ is a  proper  \Bf.
From Eqns (\ref{main00}), 
it immediately follows that \Bf\ and density are proportional, $ B =(\rho/\rho_0) B_0$, where $\rho_0$ and $B_0$ are constants taken to be density and \Bf\ in the initial state.

For $\Gamma=2$, Riemann invariants take the form
\ba &&
J_+= \log\left( \delta_\beta \delta _A^2 \right)
\nn &&
J_- = \log\left( \delta_\beta \delta _A^{-2} \right)
\label{Js}
\ea
where we   introduced Doppler factors  $\delta_A$
\be
\delta _A=  {1\over  (1-\beta_A) \gamma_A }= \sqrt{1+\beta_A\over 1-\beta_A} 
\ee
where $\beta_A$ is the local \Alfven speed.

General solution (\ref{general}) then gives
 \be
  \delta_A= C_0 \delta_\eta^{\mp 1/3},  \, \delta _\beta= \delta_\eta \delta_A ^{\pm 1}= C_0\delta_\eta^{2/3}
  \label{solution}
\ee
where $C_0$ is a constant to be determined from initial condition. Upper signs in Eq. (\ref{solution}) correspond to the forwards characteristics.

\subsubsection{Stationary initial conditions}

\begin{figure}[h!]
 \begin{center}
\includegraphics[width=.99\linewidth]{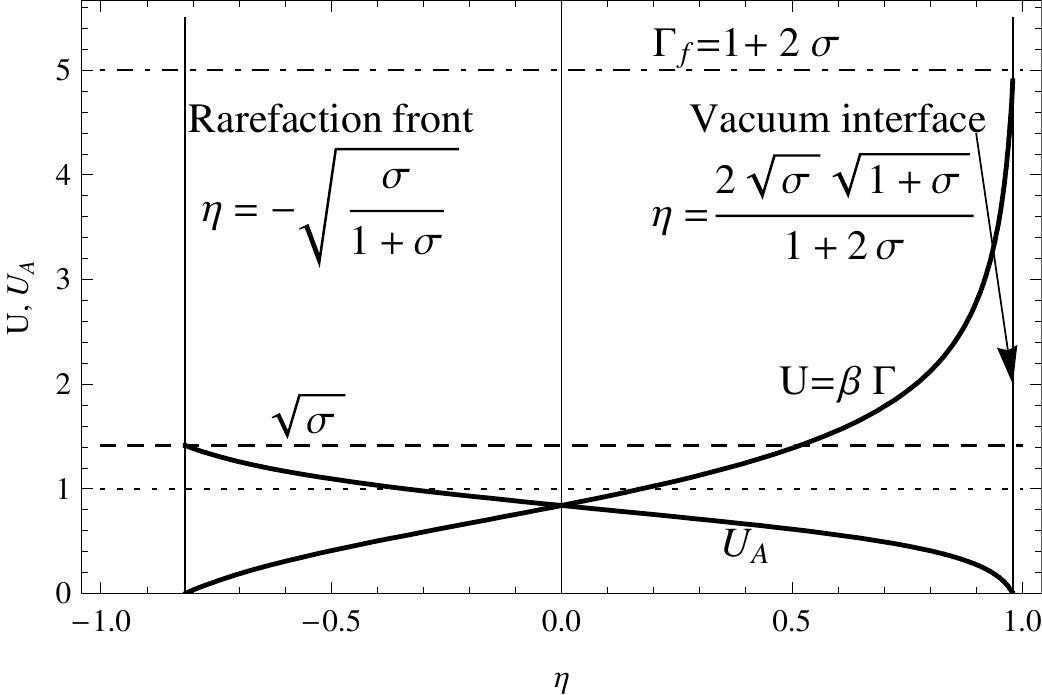}
\end{center}
\caption{ Four-velocity of the flow  and \Alfven four velocity $U_A$ for one-dimensional  self-similar expansion of magnetized gas into vacuum. For this plot $\sigma =2$. }
\label{UAU}
\end{figure}

Assuming  that initially the plasma is at rest, so that  at the front of the rarefaction wave  (where $\beta =0, \, \delta_\beta=1$) the
\Alfven velocity corresponds to the \Alfven velocity in the unperturbed medium, given by  corresponding Doppler factor $\delta_{A,0}$, we find
a simple fully analytical  solution to the problem of one dimensional relativistic expansion of magnetized gas into vacuum, see Fig. \ref{UAU}:
\be 
\delta_\beta = \delta_\eta^{2/3} \delta_{A,0}^{2/3}
, \,
\delta_A ={ \delta_{A,0}^{2/3} \over \delta_\eta^{1/3}}
\ee
The other solution, corresponding to expansion to the left is
\be
 \delta_\beta ={ \delta_\eta^{2/3} \over  \delta_{A,0}^{2/3}}, \,
  \delta_A = \delta_{A,0}^{2/3}  \delta_\eta^{1/3}.
   \ee
 These solutions give the velocity and \Bf\ as functions of the  self-similar coordinate  $\eta$ and initial magnetization $\sigma$.

The relations for the more commonly used parameters are (for the forward characteristics)
\ba &&
\gamma ={1\over 2} \left( \delta_\eta^{2/3} \delta_{A,0}^{2/3}  + {1 \over \delta_\eta^{2/3} \delta_{A,0}^{2/3} } \right)
\nn &&
\beta = {\delta_{A,0}^{4/3}\delta_\eta^{4/3} -1\over 1 +\delta_{A,0}^{4/3} \delta_\eta^{4/3}}
\nn &&
U_A = {1\over 2} \left( { \delta_{A,0}^{2/3}  \over  \delta_\eta^{1/3}} - {  \delta_\eta^{1/3} \over  \delta_{A,0}^{2/3}} \right)
\nn &&
B =\left( { \delta_{A,0}^{2/3}  \over  \delta_\eta^{1/3}} - {  \delta_\eta^{1/3} \over  \delta_{A,0}^{2/3}} \right)^2  {B_0 \over 4 \sigma}
\nn &&
\beta_A = 1-2 {  \delta_\eta^{2/3} \over  \delta_\eta^{2/3}+ \delta_{A,0}^{4/3} }
\nn &&
 \delta_{A,0} = \sqrt{ 1+   \beta_{A,0}   \over 1-   \beta_{A,0}  }, \,
  \beta_{A,0} =\sqrt{ \sigma/(1+\sigma)}
  \label{mmm}
  \ea
Plasma density at each point is $\rho= (U_A^2 /\sigma) \rho_0$ and \Bf\ is
$B=(U_A^2 /\sigma) B_0$.
We stress that these solutions are exact, no assumptions about the value of the parameter $\sigma$ were made.

In the strongly magnetized  limit $\sigma \rightarrow \infty$
we find
\ba &&
\beta = 1 - {1 \over 2^{1/3} \delta_\eta^{4/3} \sigma^{2/3}}
\nn &&
\gamma = \left( {\sigma \delta_\eta^2 \over 2 } \right)^{1/3}
\nn &&
U_A =  \left( {\sigma \over 2 \delta_\eta} \right)^{1/3} = \sqrt{\sigma \over 2 \gamma}
\label{sigmalarge}
\ea

 The front of the rarefaction wave is located where  $ \delta_\beta =1$. This gives $\delta_\eta= 1/  \delta_{A,0}$, 
 \be
 \eta_{RW} = - \sqrt{ \sigma \over 1+\sigma}
 \label{etaRF}
 \ee
 Thus the rarefaction wave propagates  into undisturbed plasma with \Alfven velocity in the unperturbed plasma $\beta_A =- \sqrt{ \sigma \over 1+\sigma}$

%          \begin{figure}[h!]
 %\begin{center}
%\includegraphics[width=.99\linewidth]{UAGamma.pdf}
%\end{center}
%\caption{ Lorentz factor $\gamma$  and \Alfven four velocity $U_A$ for one-dimensional  self-similar expansion of magnetized gas into vacuum. For this plot $\sigma =2$. }
%\label{UAGamma}
%\end{figure}

%       \begin{figure}[h!]
 %\begin{center}
%\includegraphics[width=.99\linewidth]{UAGammaSigma10.pdf}
%\end{center}
%\caption{Same as Fig. for $\sigma =10$.  }
%\label{UAGammaSigma10}
%\end{figure}

 The vacuum interface corresponds to zero magnetic pressure,  where $\delta_A=1$, Eq. (\ref{mmm}). This occurs at  $\delta_\eta = \delta_{\rm vac}= \delta_{A,0}^2$  (this corresponds to the maximum allowed value of 
 $\delta_\eta$). Vacuum interface is located at 
 \be
 \eta_{\rm vac} =  2 { \sqrt{\sigma(1+\sigma)}\over (1+2 \sigma) } \approx 
 \left\{
 \begin{array}{ll}
  1-1/(8\sigma ^2) & \mbox{ if}  \sigma \gg1 \\
  2 \sqrt{\sigma} & \mbox{ if}  \sigma \ll 1
\end  {array}
\right.
\ee
The Lorentz factor of the vacuum interface is exactly
\be
\gamma_{\rm vac} = {1  \over \sqrt{1- \eta_{\rm vac}^2}} =1 +2 \sigma
\ee
The corresponding four-velocity is
\be
U_{\rm vac} = {  \eta_{\rm vac} \gamma_{\rm vac} }= 2 \sqrt{\sigma(1+\sigma)}
\approx 
 \left\{
 \begin{array}{ll} 2 \sigma& \mbox{ if}  \sigma \gg1 \\
  2 \sqrt{\sigma} & \mbox{ if}  \sigma \ll 1
\end  {array}
\right.
\ee
Since in the limit $\sigma \ll 1$ the initial \Alfven velocity is $\beta_{A,0} = \sqrt{\sigma}$, the front expands with $v _{\rm vac} = 2 \beta_{A,0}$, in agreement with known results for polytropic fluid  \citep[expansion into vacuum with velocity $2 c_s/(\Gamma-1)$][]{1948sfsw.book.....C,GreenspanButler, ZeldovichRaizer}

%The rarefaction wave propagates into plasma at rest with \Alfven velocity in the initial state. Its front is located at 
%\ba &&
%\delta_{\rm rar}= {1\over  \delta_{A,0}}
%\nn &&
%\eta_{\rm rar} = - \sqrt{\sigma/(1+ \sigma)}
%\ea

%This gives the minimum  allowed value of $\delta$,
%\be
%\delta_{\rm rar} =1/\delta_{A,0} \approx
%\left\{
 %\begin{array}{ll}
%2 \sqrt{\sigma} & \mbox{ if}  \sigma \gg1 \\
 % 1+ \sqrt{\sigma}  & \mbox{ if}  \sigma \ll 1
%\end  {array}
%\right.
%\ee

The flow becomes sonic, $\beta_A = \beta$ always at $\eta =0$,  $\delta_\eta =1$; this is similar to the general case of non-relativistic expansion of polytropic gas into vacuum \cite[][\S 20]{Stanyukovich}. At this point the four-velocity of the flow
\be
U(\eta =0) 
\approx 
 \left\{
 \begin{array}{ll}
  (\sigma/2)^{1/3} & \mbox{ if}  \sigma \gg1 \\
  2 \sqrt{\sigma}/ 3  & \mbox{ if}  \sigma \ll 1
\end  {array}
\right.
\ee

Equation for forward and backward characteristics are
\ba &&
\left. d_t z \right|_+ = {\beta +\beta_A \over 1+ \beta \beta_A}=  {\delta_{A,0}^{8/3}\delta_\eta^{2/3} -1\over 1 +\delta_{A,0}^{8/3} \delta_\eta^{2/3}} 
\nn &&
\left. d_t z \right|_- = {\beta -\beta_A \over 1- \beta \beta_A}={z \over t}
\label{443}
\ea
The backward characteristics is a straight line, $z=\eta t$, with $\eta_{rar} < \eta< \eta_{vac} $. Equation for the forward characteristics integrates to give
\be
{t}  =  {\rm Const} {(1+\delta_\eta^2)  \over      (   \delta_{A,0}^{8/3}-\delta_\eta^{4/3} )^{3/2}}, \, \delta_\eta <  \delta_{A,0}^2
\label{444}
\ee
For a characteristics that starts at a point $t=t_0$, $z=-t_0 \sqrt{\sigma/(1+\sigma)}$, this gives
\be
{t\over t_0} = (  \delta_{A,0}^2-1) \sqrt{\delta_{A,0}^4-1} {(1+\delta_\eta^2)  \over      (   \delta_{A,0}^{8/3}-\delta_\eta^{4/3} )^{3/2}}
\label{RW2}
\ee

Thus, the forward characteristics never cross the vacuum interface ($\delta_\eta  = \delta_{A,0}^2$) and becomes parallel to it as $t \rightarrow \infty$. 
%This implies that asymptotically all the plasma acquires expansion velocity equal to the velocity of the vacuum interface, $\gamma_f= 1+2 \sigma$ and, correspondingly, zero magnetization \citep{1948sfsw.book.....C,GreenspanButler,ZeldovichRaizer}.

 The flow lines are given by
 \be
 {d z \over dt} = \beta = { \delta_{A,0}^{4/3} \delta_\eta^{4/3} - 1 \over 1+ \delta_{A,0}^{4/3} \delta_\eta^{4/3}}. 
 \label{445}
 \ee
 Which integrates to give
 \be
{t} =  {\rm Const}  {(1+\delta_\eta^2)  \over      (   \delta_{A,0}^{8/3}-\delta_\eta^{4/3} )^{3}}, \, \delta_\eta <  \delta_{A,0}^2
\label{444d}
\ee
In particular, for  a flow line starts at a point $t=t_0$, $z=-t_0 \sqrt{\sigma/(1+\sigma)}$, this gives
\be
{t\over t_0} = {(  \delta_{A,0}^2-1)^3 \over1+ \delta_{A,0}^2 } {(1+\delta_\eta^2)  \over      (   \delta_{A,0}^{8/3}-\delta_\eta^{4/3} )^{3}}
\ee
Thus, both the characteristics and the flow lines asymptote to $\delta_\eta \rightarrow  \delta_{A,0}^2$. 

      For $\sigma \gg 1$ the expansion is relativistic practically  in the whole flow.  In the bulk,  $\gamma \sim \sigma^{1/3}$ (e.g., at the sonic point $\eta =0$, 
      $\gamma= (\sigma/2)^{1/3}$).  In a narrow region near $\eta_{\rm vac}$, with thickness of the order $\Delta \eta \sim 1/ \sigma^2)$, the Lorentz factor of the flow approaches  $\gamma_{\rm vac} $ (at this  $\Delta \eta $ the Lorentz factor reaches half of its maximum values). As the flow expands, the local magnetization, $U_A^2$,  decreases. 
      
      The energy flux, 
      \be
      T_{0z}= \gamma^2 (\rho + B^2) \beta = {(\delta_\eta^{4/3} -  \delta_{A,0}^{8/3})^2 ( \delta_\eta^{8/3}\delta_{A,0}^{8/3} -1) \over  64 \delta_{A,0}^{4} \delta_\eta^{8/3} \sigma} \rho_0
      \ee
      reaches maximum at $\eta =0$. In the limit $\sigma \gg 1$ it stays nearly constant in between $\eta_{RW} $ and $\eta_{\rm vac}$ at a value  $T_{0z} \approx B_0^2/4$.  The energy density component, 
      \be
      T_{00}  = \gamma^2 (B^2 +\rho) -B^2/2= { (\delta_{A,0}^{4/3} -\delta_\eta^{2/3})^2 \left(  (\delta_{A,0}^{4/3} +\delta_\eta^{2/3} )^2 (1+\delta_{A,0}^{8/3} \delta_\eta ^{8/3}) + 8 \delta_{A,0}^{8/3} \delta_\eta^2\right)   \over 64 \delta_{A,0}^{4} \delta_{\eta} ^{8/3} \sigma } \rho_0
      \ee 
       decreases towards the vacuum interface, remaining nearly constant in between $\eta_{RW} $ and $\eta_{\rm vac}$ at the same value $T_{00} \approx B_0^2/4$.
         It may be verified by direct calculations that the total energy in the flow,   integral of $T_{00}$ from  ${\eta_{RW}} $ to ${\eta  _{\rm vac}}$, equals the total energy in the initial state between $\eta_{RW}$ and $0$, with the energy density equal to $ \rho_0 (1 +\sigma/2)$.

  \subsubsection{Moving piston: expansion into vacuum}
  
      Let us now assume that in the  undisturbed plasma, the piston, is moving with velocity $\beta_w$ towards the external medium.  The corresponding relations can be trivially obtained using Riemann invariants and imposing a condition that the piston  is moving with  velocity $\beta_w$ and  corresponding  Doppler factor $\delta_w = \sqrt{(1+\beta_w)/(1-\beta_w)}$. 
      Thus, in the previous relations we need to make a substitution $\delta_{A,0} \rightarrow \delta_{A,0}  \sqrt{\delta_w}$ (this choice assumes that the magnetized medium moves toward the unmagnetized one; in the opposite case, $\delta_{A,0} \rightarrow \delta_{A,0}  /\sqrt{\delta_w}$). 
      We  then find exact solution of relativistic Riemann problem for
      expansion into vacuum of cold  strongly magnetized plasma with magnetization parameter $\sigma$ and moving with velocity $\beta_w$: 
       \ba &&
\delta_\beta = \delta_\eta^{2/3} \delta_{A,0}^{2/3} \delta_w ^{1/3}
\nn &&
\delta_A ={ \delta_{A,0}^{2/3}  \delta_w ^{1/3} \over \delta_\eta^{1/3}}
\nn &&
\gamma ={1\over 2} \left( \delta_\eta^{2/3} \delta_{A,0}^{2/3}   \delta_w ^{1/3} + {1 \over \delta_\eta^{2/3} \delta_{A,0}^{2/3}  \delta_w ^{1/3}} \right)
\nn &&
\beta = {\delta_{A,0}^{4/3}\delta_\eta^{4/3}  \delta_w ^{2/3} -1\over 1 +\delta_{A,0}^{4/3} \delta_\eta^{4/3} \delta_w ^{2/3}}
\nn &&
U_A = {1\over 2} \left( { \delta_{A,0}^{2/3}  \delta_w ^{1/3} \over  \delta_\eta^{1/3}} - {  \delta_\eta^{1/3} \over  \delta_{A,0}^{2/3} \delta_w ^{1/3}} \right)
\nn &&
\beta_A = 1-2 {  \delta_\eta^{2/3} \over  \delta_\eta^{2/3}+ \delta_{A,0}^{4/3}  \delta_w ^{2/3}}
\label{main}
\ea
(In case of a boost away from the contact, we should substitute  $  \delta_w ^{1/3} \rightarrow   \delta_w ^{-1/3}$, in accordance with the Lorentz transformation of the Doppler factor.)

The vacuum interface is moving with $\delta _\eta=  \delta_{A,0}^{2} \delta_w$, which in the limit $\sigma, \, \gamma_w \gg1$ gives
$\gamma_{vac} = 2 \gamma_w (1+ 2 \sigma)$. 

The front of the rarefaction wave is located at $\delta_\beta = \delta_w$, which  gives
\be
\delta_\eta = {  \delta_w \over \delta_{A,0}} \approx {\gamma_w \over \sqrt{\sigma}}
\ee
The  front of the rarefaction wave is stationary  (located at $\eta =0, \, \delta_\eta =1$) when the flow moves sonically, with the \Alfven velocity of the undisturbed plasma,  $\beta_w = \beta_{A,0}$. 
 We stress that these solutions are exact, no assumptions about the value of the parameter $\sigma$ and the  velocity $\beta_w$ were made. 
 
  \subsubsection{Reflection of the rarefaction wave from the wall}
  
\begin{figure}[h!] 
 \begin{center}
\includegraphics[width=.49\linewidth]{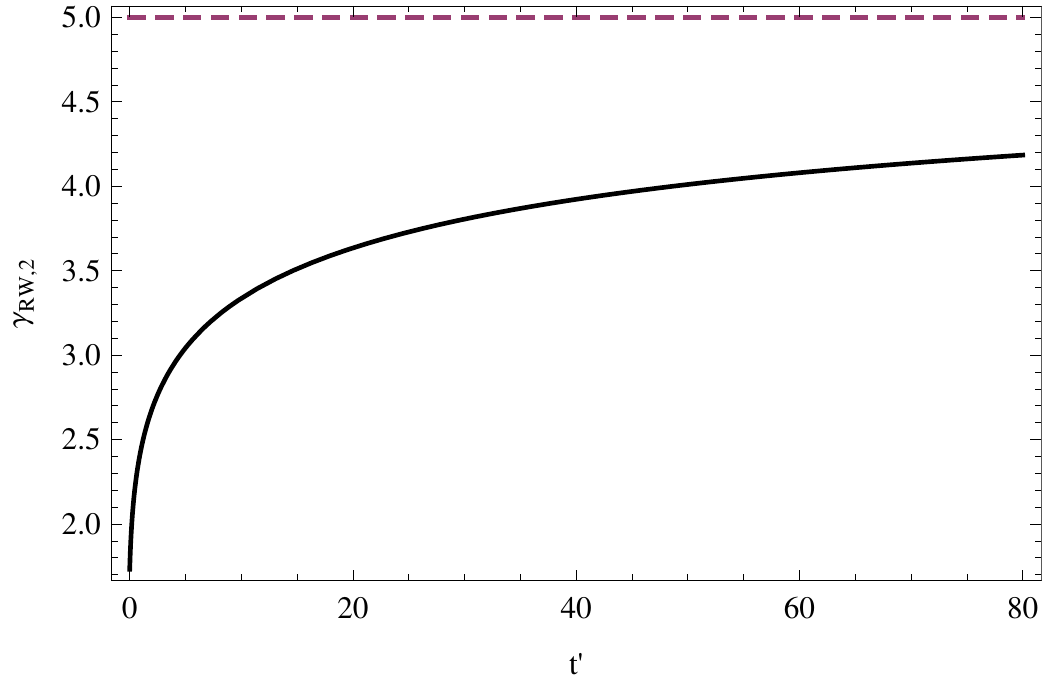}
\includegraphics[width=.49\linewidth]{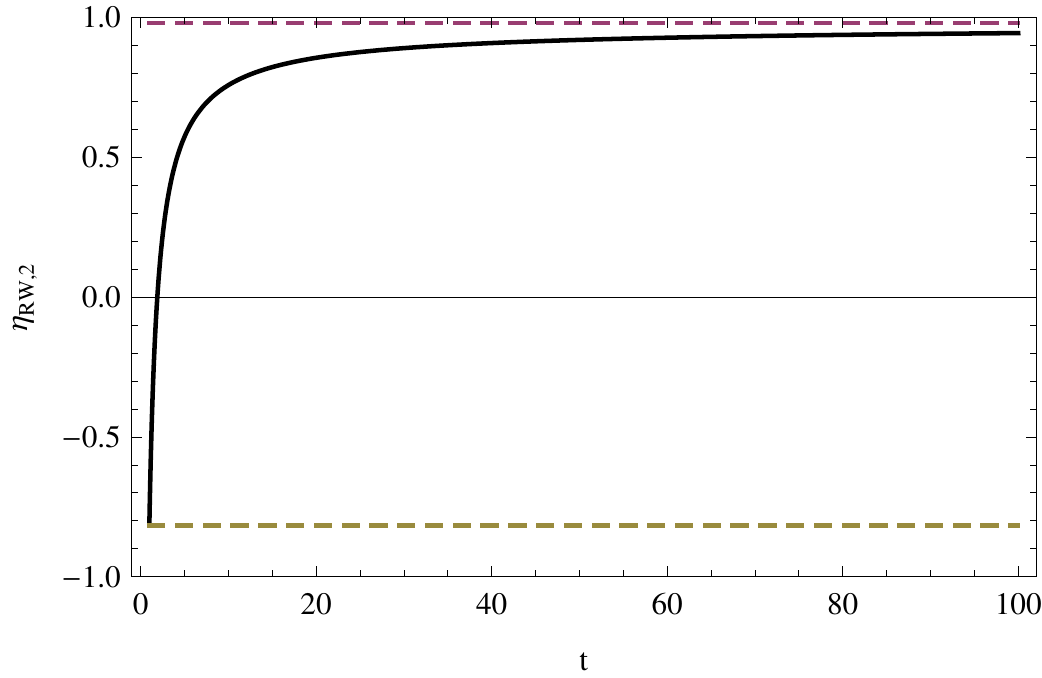}
\end{center}
\caption{{\it Left Panel}: Lorentz factor of the secondary rarefaction wave as function of  $t'$, time since reflection. The secondary RW propagates in the flow disturbed by the primary RW. In the limit $t' \gg 8 \sigma^2 t_0$, the secondary RW   reaches the terminal Lorentz factor of the vacuum interface $\gamma_{\rm vac}  = 1+ 2\sigma$ (dashed line). The plot is for $\sigma=2$.   The limiting value of the Lorentz factor  is reached very slowly, with $\gamma_{RW,2}  = \sqrt{2} \sigma^{2/3} (t/t_0)^{1/6}$ for $t \ll  8 \sigma^2 t_0$.
{\it Right Panel}: motion of the secondary RW in self-similar coordinates $\eta = z/t$. At time $t_0= 1$, secondary RW is launched from $z= - \sqrt{\sigma/(1+\sigma)} t_0$, corresponding to  $\eta = - \sqrt{\sigma/(1+\sigma)}$  (lower dashed line)  in the forwards direction. At $ t \rightarrow \infty$ the front of the RW approaches the  vacuum interface $\eta =  2 \sqrt{ \sigma(1+\sigma)}/(1+ 2 \sigma)$ (upper dashed line).  }
\label{gammaRW2}
\end{figure}

  Let us now assume that the initial state with $\beta_w=0$ occupies  a limited region of space $- L < z < 0 $, with impenetrable wall at $z=-L$. As the rarefaction wave reaches the wall, it  will be reflected, creating  a secondary RW.  The initial RW propagates to the left, while the secondary RW propagates to the right into the region disturbed by the first RW. Let's denote $\beta_{RW,2}$ the velocity of  the secondary RW. Its front propagates along the forward characteristics with velocity given by Eq. (\ref{RW2}), Fig. \ref{gammaRW2}.
    
  In this case the expansion is non-self-similar, since  there is a typical scale in the problem $L$. 
  In addition, even at asymptotically long times after the beginning of the flow,  at times $t \gg L/c$, the  expansion of 
        a magnetic shell  is not self-similar, since
        conservation of mass (and magnetic flux) results in  different scaling of magnetic ($\propto B^2$) and rest mass energy densities.
        
        Still, it might be possible to find exact solution for the
  expansion of a magnetized layer into plasma following the corresponding hydrodynamic  approach \citep{BelenkijLandau}. (One can always replace  the nonlinear problem of  one-dimensional hydrodynamic evolution with a linear equation using \cite{Khalatnikov} transform). We leave this problem for future; here we consider the motion of the front of the secondary RW. 
     
   Let us discuss the limiting cases  of flow evolution long after the reflection of the secondary RW from the wall, $t\gg t_0$. In the high magnetization limit, $\delta_{A,0} = 2 \sqrt{\sigma} \gg1$, and for large Lorentz factors of the the secondary RW, $\beta_{RW,2} \sim 1-1/(2 \gamma_{RW,2}^2)$, Eq. (\ref{RW2}) gives
   \be
   {1 \over 2 \gamma_{RW,2}^2} = { \left( t_0 + t/(2 \gamma_{RW,2}^2) \right)^{1/3} \over 4 t^{1/3} \sigma^{4/3}}
   \ee
   For a long time, when $t \leq 2 t_0 \gamma_{RW,2}^2$,  the rarefaction accelerates only slowly,
   \be 
   \gamma_{RW,2}  = \sqrt{2} \sigma^{2/3} (t/t_0)^{1/6}, \, \mbox{if} \, t \leq  8 t_0 \sigma^2 
   \label{34}
   \ee
   For $t > 2 t_0 \gamma_{RW,2}^2$, which using Eq. (\ref{34}) implies $t> 8 t_0 \sigma^2$, the secondary RW  reaches the terminal Lorentz factor
   $\gamma_{RW,2}  \approx 2 \sigma$, the Lorentz factor of the vacuum interface.        In self-similar coordinates, the front of the secondary RW moves from $\eta = -   \sqrt{\sigma/(1+\sigma)}$ to $\eta=  2 \sqrt{ \sigma(1+\sigma)}/(1+ 2 \sigma)$, Fig. \ref{gammaRW2}.
            Note, that even in the case when the second rarefaction wave approaches the vacuum interface, the expansion is still generically 
       non-self-similar since  there are two conserved quantities, energy and magnetic flux,  that scale differently with radius. 
       
       The velocity of plasma in front of  the reflected rarefaction waves is given by Eq. (\ref{445}). The flow initially accelerates 
      \be
      \gamma = (t/t_0)^{1/3}  \sigma^{1/3}
      \ee
       and reaches terminal Lorentz factor $\gamma = 1+ 2 \sigma$ at times $t \geq  8 t_0 \sigma^2$ (see also \cite{2011MNRAS.411.1323G}). As the secondary rarefaction wave catches up with the flow, the flow is decelerated. Thus, most of the acceleration occurs in a regime when the flow is causally disconnected from the wall (contrary to the claim in Ref.  \cite{2010arXiv1004.0959G}, version 1). 
       
       \section{One-dimensional expansion of magnetic piston into plasma}
       
       \subsection{Stationary initial conditions}
       
              The solution for the expansion of magnetized plasma into vacuum 
     derived in \S  \ref{Simple} can be used to consider a decay of a contact discontinuity between magnetized plasma with density $\rho_0$ and magnetization parameter $\sigma $ and a cold plasma with density $\rho_{\rm ex}$. For clarity, we refer below to the magnetized component as a piston, and the non-magnetized plasma as the  external medium.
     
   For cold external plasma,   we expect the  formation of four regions: undisturbed external  plasma, shocked  external plasma, expanding magnetic piston and 
     undisturbed magnetized plasma.   
     These regions are separated by 
     the    forward shock, a contact discontinuity and an expansion wave.  We expect that 
     the dynamics of the system will be self-similar. 
     
     The solution for simple waves derived in \S  \ref{Simple} can be used to describe the expansion wave part of the flow. For  non-zero external density, the simple waves will terminate at a contact discontinuity (CD), located at a particular value of the self-similar variable $\eta_{CD} < \eta_{\rm vac} $, so that the magnetic pressure of the expansion wave balances the thermal pressure of the shocked medium, which, in turn, depends of the velocity of the CD and external density $\rho_{\rm ex}$.

      Let $\beta_1$ and $\beta_2$ be velocities of the unshocked  and shocked plasma in the frame of the forward shock. 
      The shock jump conditions require \citep[assuming $p_1=0$,][]{LLIV}
      \ba &&
      \beta_1^2 = {\epsilon _2 p_2\over (\epsilon _2-\rho_{\rm ex})(\rho_{\rm ex}+p_2) }
      \nn &&
      \beta_2^2 = {p_2 (p_2+\rho_{\rm ex}) \over \epsilon _2 (\epsilon _2- \rho_{\rm ex})}
      \label{v1v2}
      \ea
      where $\epsilon_2$ and $p_2$ are energy density and pressure in the shocked medium and  we assumed that the external medium is cold, $p_{\rm ex}=0,\, \epsilon_{\rm ex}=\rho_{\rm ex}$. 
      
      In the frame of the CD (CD is stationary with respect to downstream plasma), the shock 
has  velocity $v_2$, while  the incoming flow has velocity $ v_{12}$. In the observer' s frame, the CD have velocity $ v_{CD}$, while the shock has
      \be 
      \beta_{FS} = {\beta_{CD} +\beta_2 \over 1+ \beta_{CD} \beta_2}
      \label{vFS}
      \ee
      Since the upstream  plasma is at rest in the observer's frame
      \be
      \beta_{CD}= v_{12}
        \label{vCD}
      \ee

      Equations (\ref{vCD}), an 
   equation of state $\epsilon _2(p_2)$, a condition of force balance on CD 
   \be
   p_2= B^2/2 = U_A(\eta = \eta_{CD}) ^4 \rho_0 /(2 \sigma)
   \label{p2}
   \ee
   and shock jump conditions \citep{1948PhRv...74..328T}  
   \be
 \left(  {{\cal E}_1 + \rho_1+ p_1\over \rho_{\rm ex}} \right) ^2 =  \left(  {{\cal E}_2 +\rho_2+ p_2 \over \rho_{2}} \right) ^2 - (p_2-p_1) 
 \left(  {{\cal E}_1 +\rho_1+ p_1\over \rho_{\rm ex}^2} +  {{\cal E}_2 +\rho_2+
  p_2\over \rho_{2}^2} \right)
 \label{Taub}
 \ee
    constitute an equation on $\eta_{CD}$, the location of the contact discontinuity. (In Eq. (\ref{Taub}) indices  $1$ and $2$ refer to unshocked
    and shocked external media correspondingly, $\rho_1\equiv \rho_{\rm ex}$.)
    
    For cold external plasma, $p_1=0$ and assuming adiabatic index of $\Gamma=4/3$ (so that ${\cal E}_2= 3p_2 +\rho_2$), the shock jump condition 
    gives
    \be
    \rho_2 = {1\over 2} \left(  7 \rho_1 +\sqrt{ \rho_{\rm ex} ( 48 p_2 + 49  \rho_{\rm ex}) }\right)
    \ee
    
\begin{figure}[h!] 
 \begin{center}
\includegraphics[width=.49\linewidth]{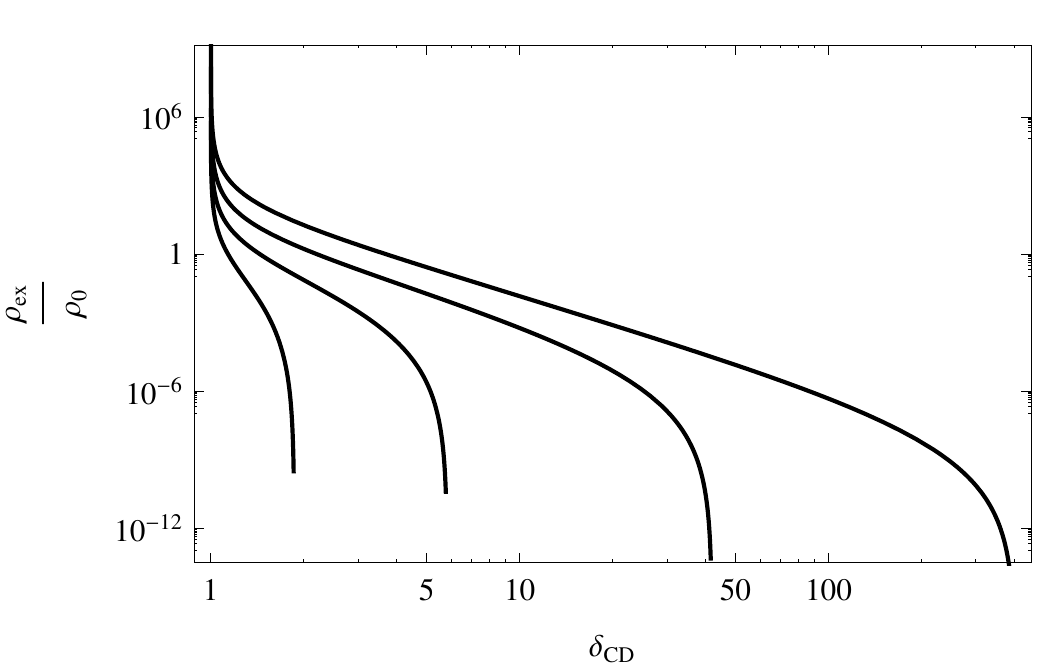}
\includegraphics[width=.49\linewidth]{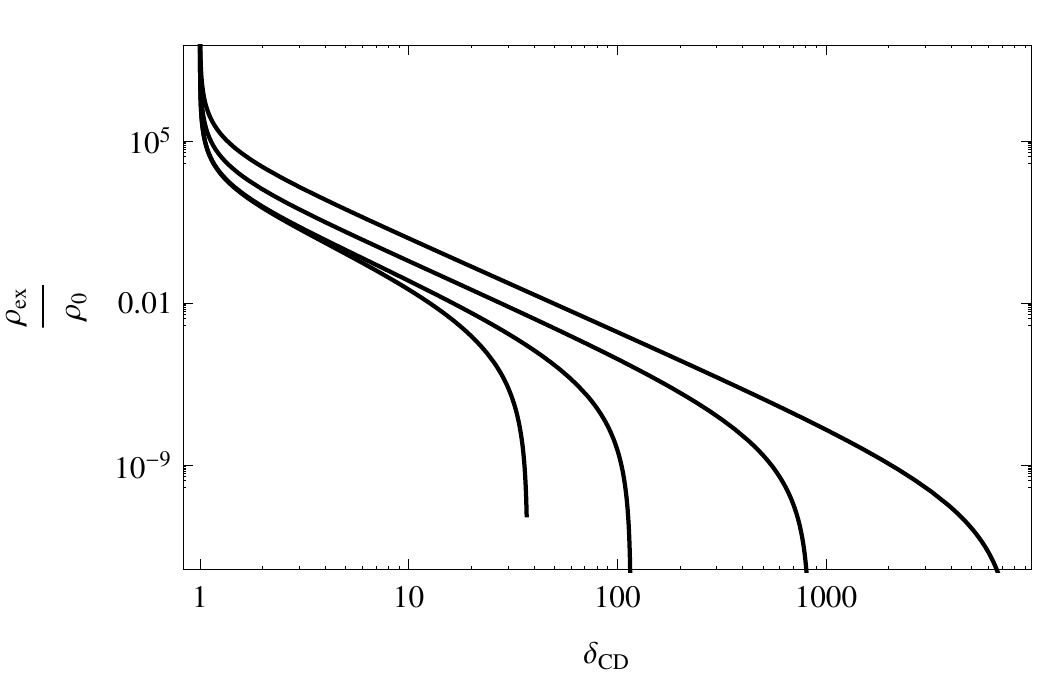}
\end{center}
\caption{Ratio of external density to the piston density $\rho_{\rm ex}/\rho_0$ as a function of the 
Doppler factor of the contact discontinuity  $\delta_{CD}$ for one-dimensional expansion of strongly magnetized plasma with magnetization $\sigma$, 
$\sigma= 100, 10, 1, 0.1$, top to bottom curves.  {\it Left Panel}: Stationary initial conditions.  {\it Right panel}: piston moving with $\gamma_w=10$.  Maximum values of $\delta_{CD}= \delta_{A,0}^2$ (left panel) and $\delta_{CD}= \delta_{A,0}^2 \delta_w$ (right panel) correspond to zero external density, a vacuum interface.
}
\label{deltaCD}
\end{figure}

    Using continuity equations (\ref{v1v2}-\ref{vCD}), solutions for the expansion flow  given by (\ref{mmm}) and shock jump condition (\ref{Taub}) we find
    \be
    {\rho_{\rm ex}  \over \rho_{0}} =-  { 3 (\delta_{A,0}^2 - \delta _{\beta, CD})^4 \over
    16 \sigma  \delta_{A,0}^4 (1 -  \delta _{\beta, CD}^2) ( 2+ 3 \delta_{\beta, CD} + 2 \delta _{\beta, CD}^2) }
    \label{betaCD}
    \ee
    Eq.  (\ref{betaCD}) determines the Doppler factor of the contact discontinuity for plasma obeying equation of state with polytropic index $\Gamma =2$, 
    having initial \Alfven Doppler factor $\delta_{A,0}$
     expanding into a medium with density $\rho_{\rm ex} $; the external medium is unmagnetized and obeys  an equation of state with polytropic index $\Gamma =4/3$. Eq.  (\ref{betaCD}) is relativistically exact, no assumption about values of magnetization or external density were made, see Fig. \ref{deltaCD}. Note, that  it correctly reproduces expansion into vacuum, 
     $\rho_{\rm ex}=0, \, \delta _{vac} = \delta_{A,0}^2$. Thus, shock jump conditions (\ref{Taub}) are also applicable to the plasma-vacuum interface.
     
In the limit of relativistically strong shock, when the post-shock pressure is much larger than the post-shock density, relation
(\ref{betaCD}) gives
\be
\gamma_{CD} = \left( {3 \over 32} {\rho_0 \over \rho_{\rm ex}} \sigma \right)^{1/4}= \left( {3 \over 32} {B_0^2 \over \rho_{\rm ex}}  \right)^{1/4}
\label{gammaCD}
\ee
   In this limit the  velocity of the forward shock in the observer frame,
$\beta_{FS} =( \beta_{CD} +\beta_2)/(1+\beta_{CD}  \beta_2)$, is  
\be
\gamma_{FS} =\sqrt{2}  \gamma_{CD} = \left( {3 \over 8} {\rho_0 \over \rho_{\rm ex}} \sigma \right)^{1/4} =
 \left( {3 \over 8} {B_0^2 \over \rho_{\rm ex}} \right)^{1/4}
 \label{gammaFS}
\ee

Heuristically, the expressions (\ref{gammaCD},\ref{gammaFS})  can be obtained as a pressure balance at time $t=0+$, when the contact instantaneously reach a Lorentz factor 
      $ \gamma_{CD} $;   the magnetic pressure in the plasma frame then is  $\sim B^2/\gamma_{CD}^2$, and the kinetic pressure of the shocked external medium, $\sim \rho_{\rm ex} \gamma_{CD}^2$. 
     
      Note, that  in the strong FS limit, the Lorentz factor of the CD, and of the forward shock, Eqns (\ref{gammaCD},\ref{gammaFS}) are {\it independent} of $\sigma$, the  composition of the driver.
     %,  contrary to the claims by \cite{Mizuno}. 
     Only the value of total pressure in the piston matters ($\propto B_0^2$ in our case). 
Composition of the driver becomes important only for weak, non-relativistic  FS, when the Lorentz factor of the CD (\ref{gammaCD}) approaches the limit of expansion into vacuum. $\gamma_{vac} =  1+2 \sigma$. 
This occurs for 
\be
\rho_{\rm ex} < {3 \over 256} {1\over \sigma^3} \rho_0
\label{rhoex}
\ee
In this limit
    of small  external density,  $\rho_{\rm ex} \rightarrow 0$, we find
      \be
      \gamma_{CD} = 1 + \left( 2- \left( {14336 \over 3} {\rho_{\rm ex}  \over \rho_0 }  \right)^{1/4} \right) \sigma 
   \ee

   \subsection{Expansion into medium:  moving piston}
   
   For moving piston, the system becomes somewhat more complicated.
    Similarly to the stationary case,   we expect the  formation of four regions: undisturbed external  plasma, shocked  external plasma, undisturbed magnetized plasma;  but the expanding magnetic piston  may now turn into region  of shocked piston  materiel.      These regions are separated by 
     the    forward shock, a contact discontinuity and an expansion wave or the reverse shock front.  
          
     The cases, of the expansion wave and the reverse shock have somewhat different dynamics.
     In case of the expansion wave, 
     the solution for simple waves derived in \S  \ref{Simple} can be used to describe the expansion wave part of the flow.           Alternatively, if RS shock is launched, which heats the piston material, the pressure on CD is determined by the sum of the magnetic pressure of \Bf\ compressed at the RS and the kinetic pressure of shocked particles.

\subsubsection{Forward shock and rarefaction wave}

In case of  magnetic piston moving with  initial Doppler factor $\delta_w$  into cold medium with density $\rho_{\rm ex}$,  the CD  is moving with the Doppler factor $ \delta _{\beta, CD,w}$ determined from  
 \be
     {\rho_{\rm ex}  \over \rho_{0}} =-  { 3 (\delta_{A,0}^2  \delta_w - \delta _{\beta, CD,w})^4 \over
    16 \sigma  \delta_{A,0}^4  \delta_w^2 (1 -  \delta _{\beta, CD,w}^2) ( 2+ 3 \delta_{\beta, CD,w} + 2 \delta _{\beta, CD,w}^2) }
    \label{betaCDw}
    \ee
   For strong forward shocks, the Lorentz factors of the CD and the FS are
 \ba &&
\gamma_{CD,w} =
\left({3 B_0^2 \gamma_w^2 \over 8 \rho_{\rm ex}} \right)^{1/4} 
\nn &&
\gamma_{FS,w} =
\left({3 B_0^2 \gamma_w^2 \over 2 \rho_{\rm ex}} \right)^{1/4} 
\label{gammaCD1}
\ea
(see Fig. \ref{deltaCD}).

Composition of the driver becomes important only for weak FS, when the Lorentz factor of the CD (\ref{gammaCD}) approaches the limit of expansion into vacuum. $\Gamma =  2 \gamma_w (1+2 \sigma)$. 
This occurs for 
\be
\rho_{\rm ex} < {3 \over 1024} {1\over \gamma_w ^2 \sigma^3} \rho_0
\label{rhoex1}
\ee
The  front of the rarefaction wave is stationary  (located at $\eta =0, \, \delta_\eta =1$ when the flow moves sonically, with the \Alfven velocity of the undisturbed plasma,  $\beta_w = \beta_{A,0}$. For higher $\beta_w$ the front of the RW is advected forward. 

\subsubsection{Formation of a RS: supersonic motion of ejecta in the CD frame}

For sufficiently high velocity $\beta_w$, the location of the RW ,  Eq. (\ref{etaRF}),  coincides with the location of the FS
 $\delta_{\eta,FS} \approx 2 \gamma_{CD}$. This occurs for 
 \be
 \gamma_w > 2 \gamma_{CD} \sqrt{\sigma}, 
 \label{RS3}
 \ee that is, when the wind velocity in the frame of the CD is supersonic, an obvious condition.
In terms of initial magnetization and the ratio of densities, the RS forms when 
\be
\gamma_{ w} > \gamma_{ w, crit} =   \sqrt{{3 \over 8} {\rho_{0} \over \rho_{\rm ex}} \sigma } \delta_{A,0}^2 
=
\left\{
\begin{array}{cc}
 \sqrt{{3 \over 8} {\rho_{0} \over \rho_{\rm ex}} \sigma }, & \sigma \ll 1
 \\
 \sqrt{ 6 {\rho_{0} \over \rho_{\rm ex}} } \sigma^{3/2}, & \sigma \gg 1
\end{array}
\right.
\label{RS1}
\ee
Relation (\ref{RS1}) assume strong FS and high initial Lorentz factor $\gamma_w \gg 1$. 

\subsubsection{Formation of a RS: subsonic motion of piston in the CD frame}

When motion of the piston in the frame of the CD is subsonic,  yet the piston is moving faster than the CD, $\gamma_{CD} < \gamma_w < 2 \gamma_{CD} \sqrt{\sigma}$, the flow is decelerated by a compression wave. For one-dimensional motion the compression waves are unstable to formation
of shocks  \citep{LLIV}, so that  {\it the   reverse shock will form for $\gamma_w > \gamma_{CD} $, and not at the condition (\ref{RS3})}. 
As long as  $ \gamma_w < 2 \gamma_{CD} \sqrt{\sigma}$, the  reverse shock is weak, 
in this range the RS may not form, if a more complicated flow patters are allowed. 
For $\gamma_w > 2 \gamma_{CD} \sqrt{\sigma}$, RS becomes strong. (We define strong shocks as the shocks in which the upstream four-velocity in the frame of the shock is much larger than the upstream \Alfven velocity).

\subsection{Reverse shock}

For sufficiently fast initial velocity, satisfying , $\gamma_{ w} \gg \gamma_{ w, crit}$, highly magnetized, $\sigma \gg 1$, reverse shock forms. In this section we consider the dynamics of double-shock structures.
For convenience we will make two approximations; first, we assume the the forward shock is strong and unmagnetized; second, we assume that the reverse shock is strong as well, $\gamma_w \gg \gamma_{ w, crit}$. 
We have to solve simultaneously for two shock jump condition and flow continuity at the CD.

In the frame of the CD, the FS is moving with $\beta_{FS}' =1/3$, while reverse shock is moving with  \citep{kc84}
\be 
\beta_{RS} ^{\prime , 2} =  
\frac{1+10 \sigma+  8 \sigma ^2+ (1+2 \sigma ) \sqrt{ 1+16 \sigma+ 16 \sigma ^2}}{17+ 26 \sigma +8
   \sigma ^2+ (1+2 \sigma ) \sqrt{  1+16 \sigma+ 16 \sigma ^2}}
   \ee
In the frame of the shock, the post shock kinetic pressures are  \citep{kc84}
\be
{p_{2, kin} \over \rho_1 u_1^2} =  \frac{\sigma  \left(1-\frac{\gamma _2}{u_2}\right)+1}{4 u_2 \gamma_2 }
\ee
while \Bf\ satisfies
\be
B_1 \gamma_1 \beta_1 = B_2 \gamma_2 \beta_2
\ee
where indices $1$ and $2$ refer to quantities measured in the upstream and downstream {\it in the frame of the shock}.

If the CD is moving with velocity  $\beta_{CD}$ and the initial velocity is $\beta_w$, the velocity of the incoming plasma in the frame of the RS is
\be
v_{1,RS}'={ (1- \beta_{CD} \beta_w ) \beta_{RS} ' + \beta_w -\beta_{CD} \over
1- \beta_{CD}  (\beta_w +  \beta_{RS} ' ) + \beta_w\beta_{RS} '  }
\ee

Using the above relations, the pressure behind the FS is
\be
p_{2,FS}= {(1+3 \beta_{CD})^2 \over 12 (1-\beta_{CD}^2)} \rho_{\rm ex} \approx
{4 \over 3} \gamma_{CD}^2  \rho_{\rm ex}
\label{pFS}
\ee
Kinetic pressure behind the RS is
\be
p_{2,kin, RS} =\frac{ \gamma _{  {CD}}^2 \gamma _w^2 \left(\sigma  \beta _{  {RS}}+\beta
   _{  {RS}}-\sigma \right) \left(-\beta _{  {CD}} \left(\beta _{  {RS}} \beta
   _w+1\right)+\beta _{  {RS}}+\beta _w\right){}^2}{4 \beta _{  {RS}}^2}   \rho_0 \approx
   {1\over 8} {\gamma_w ^2 \over \gamma_{CD}^2} \rho_0\, \mbox{ if }  \sigma \gg 1
   \ee
   Magnetic pressure
is
\be
p_{2,mag, RS} =
\frac{1}{2} \gamma _{  {CD}}^2 \gamma _w^2 \left(-\beta _{  {CD}} \left(\beta
   _{  {RS}}+\beta _w\right)+\beta _{  {RS}} \beta _w+1\right){}^2
    \approx
   {1\over 2} {\gamma_w ^2 \over \gamma_{CD}^2} B_0^2 \, \mbox{ if }  \sigma \gg 1
   \label{pRS}
   \ee
   Balancing the pressures (\ref{pFS})  and  (\ref{pRS}), we find
   \be
   \gamma_{CD} =  \left( {3 \over 32}  (1+ 4 \sigma) \gamma_w ^2 { \rho_0 \over \rho_{\rm ex}}\right)^{1/4}
   \approx  \left( {3 \over 8}    \gamma_w^ 2 { B_0^2  \over \rho_{\rm ex}}  \right)^{1/4}\, \mbox{ if }  \sigma \gg 1
   \ee
    The Lorentz factor of the CD in case of reverse shock coincides with the one obtained in case of rarefaction wave, Eq. (\ref{gammaCD1}). 
   Thus, the forward shock is not influenced by the rarefaction wave-shock wave transition. 
   We stress, again, that in the strong shock limit and  for $\sigma \gg1 $,  the motion of the CD and the forward shock are  independent of the composition of the piston (its density $\rho_0$).

 If we express $\gamma_{CD}$ as a function of the total luminosity $L/S = \gamma_w^2  (B_0^2 +\rho_0) $ (assuming $\beta_w \sim 1$), where $S$ is the cross-section of the flow, we find
 \be
 \gamma_{CD} = \left( {3 \over 32} { L \over S \rho_1} { 1+4 \sigma \over 1+ \sigma}  \right)^{1/4}
 \ee
 It is only weakly dependent on $\sigma$.
 
 The RS is stationary in the observer frame when
$\beta_{CD}= \beta_{RS} ^{\prime}$, which for highly magnetized medium   $\sigma \gg 1$ (below, in this section, all relations are given for $\sigma \gg 1$) gives
\be
\gamma_w = 
\sqrt{ {8 \over 3}  { \rho_{\rm ex}  \over \rho_0} \sigma}
\label{gammawST}
\ee
In this case $ \gamma_{CD} = \sigma^{1/2}$, which
 can be understood as the RS front is receding at almost with $ u_A =
 \sigma^{1/2}$ in the CD frame.
For higher $\gamma_w$ the RS is advected towards the interface, while for smaller $\gamma_w$ it propagates in the opposite direction.
Also, when $\gamma_{CD}=\gamma_w$, $\gamma_w= \sqrt{(3/8) (\rho_0/\rho_{\rm ex} )\sigma}$, there is no reverse shock or rarefaction wave; the motion of the wind matches exactly the motion of the CD.

For values of $\gamma_w$ sufficiently different from (\ref{gammawST}), the 
 Lorentz factor of the reverse shock in the frame  of stationary external medium is then
\be
\gamma_{RS}= {1\over 2} \left( { \sqrt{\sigma} \over \gamma_{CD} } + {\gamma_{CD}  \over \sqrt{\sigma} } \right)  =\left\{
\begin{array}{ll}
 \left( {3 \over 128} \right)^{1/4} { \sqrt{\gamma_w}   \over  \sigma^{1/4}} \left( {\rho_{0} \over \rho_{\rm ex}}\right)^{1/4} & \mbox{if $\gamma_{CD}   \gg  \sqrt{\sigma}$, RS moving forward}\\
 { \sigma^{1/4} \over 6^{1/4} \sqrt{\gamma_w} }  \left( {\rho_{\rm ex} \over \rho_{0}}\right)^{1/4} & \mbox{if $\gamma_{CD}   \ll  \sqrt{\sigma}$,RS moving backward}
\end{array}
\right.
\label{1}
\ee
Recall, that relations  (\ref{1}) are applicable only if condition (\ref{RS1}) is satisfied.

 %The 
 %Lorentz factor
 % of the unshocked ejecta in the frame of the RS,
%\be
%\gamma_{w,RS}= { \gamma_w \sqrt{\sigma} \over \gamma_{CD}} =
%\left( { 8 \over 3} \right)^{1/4}
 %{ \sqrt{\gamma_w} \sigma^{1/4}  } 
%\left( {\rho_{\rm ex} \over \rho_{0}}\right)^{1/4}
%\ee
%The relative Lorentz factor of the shocked ejecta with respect to the unshocked ejecta is
%\be
%\bar{\gamma}_{CD} = \gamma_w/(2 \gamma_{CD} )= {1\over 6^{1/4} } { \sqrt{\gamma_w} \over \sigma^{1/4}  }  \left( {\rho_{\rm ex} \over \rho_{0}}\right)^{1/4} 
%\ee
%\cite[\cf][]{Sari95}. Condition $\bar{\gamma}_{CD}  \gg 1$ implies a strong reverse shock. 

\subsection{Upshot: expansion of magnetized plasma into medium}

\begin{figure}[h!]
 \begin{center}
\includegraphics[width=1.1\linewidth]{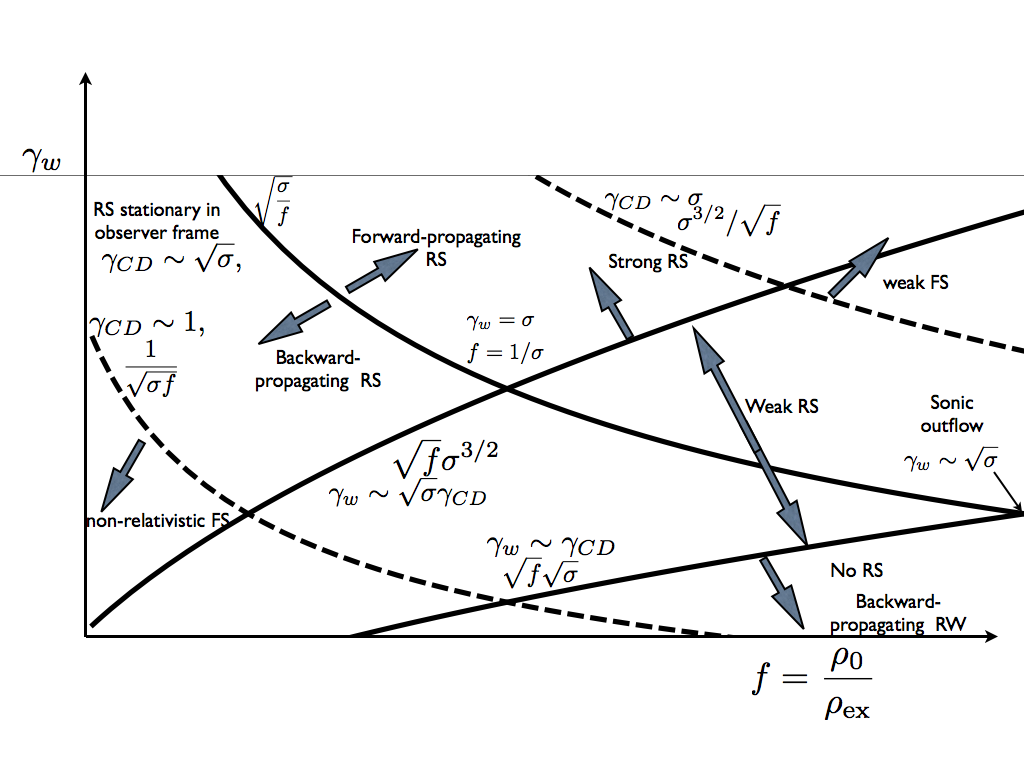}
\end{center}
\vskip -.5 truein
\caption{Cartoon of the flow dynamics as a function of two parameters, $\gamma_w$, the initial Lorentz factor for the piston and the ratio of piston density to external density $f=  {\rho_{0} \over \rho_{\rm ex}}$. High magnetization is assumed, $\sigma \gg 1$. Solid lines  are related to the reverse shock, dashed lines are  related to the forwards shock. The RS form for $\gamma_w > \gamma_{CD}$, which is $ \sqrt{f} \sqrt{\sigma} $ line. Below this line rarefaction wave propagates into piston. The RS shock is weak for  $\gamma_{CD} < \gamma_w < 2 \gamma_{CD} \sqrt{\sigma}$, the latter corresponds to  $ \sqrt{f} \sigma^{3/2}$ line. In the frame of the undisturbed plasma the revers shock propagates in the forward direction  for $  \gamma_{CD}> \sqrt{\sigma}$, which corresponds to $\gamma_w > \sqrt{f/\sigma}$. The forward shock becomes non-relativistic for $\gamma_{CD} \sim 1 $, which is the $1/\sqrt{\sigma f}$ curve. The  forward shock becomes weak (so that the post-shock temperature is non-relativistic) for $\gamma_{CD} \sim \sigma$, which is  $\sigma^{3/2} /\sqrt{f}$ line. For weak FS, the expansion proceeds nearly as into vacuum.   (Since relativistic motion is assumed, the lowest part of the plot,  below the intersection of lines $\gamma_{CD}= 1$ and $\gamma_w = \gamma_{CD}$,  is outside of the applicability region.)}
\label{Conditions}
\end{figure}

In Fig. \ref{Conditions} we qualitatively outline the dynamics of the flow expansion in a medium. As one can see, the dynamics is rich in details and is much more complicated than for pure hydrodynamical expansion. Qualitatively, as the Lorentz factor of the piston $\gamma_w$ increases for given density ration $f$, the rarefaction wave turns first into a weak RS and then into a strong RS. (In non-one-dimensional flows weak reverse shocks can be avoided.) For extremely low  external density, the expansion proceeds similarly to the case of expansion into vacuum. As the external density increases,  first, the forward shock becomes strong. For higher $\rho_{ex}$ RS becomes backward propagating. 

   \section{Three dimensional expansion into vacuum. }
   
 In case of three-dimensional expansion of plasma carrying  toroidal field, we expect that the flow will not be purely radial, as magnetic hoop stresses and magnetic pressure will, generally, induce lateral  motion (in $\theta$ direction) in addition to radial expansion. Bur for strongly relativistic motion, corresponding to high magnetization parameter, the $\theta$ component of the velocity will be much smaller, by a factor $\sim \gamma^2$ than the radial component, so that the motion will be approximately conical,   depending only  on time and radial coordinate.
 
For the  three  dimensional expansion, the  governing equations are generically non-self-similar: there are two conserved quantities, energy and magnetic flux that scale differently with radius. Self-similarity can still be achieved  in a narrow region, e.g. near the surface of the bubble, where different radial scaling  of energy  and magnetic flux can be neglected.

 Similarly to the 1-D case,  in case of three dimensional expansion of magnetized plasma into vacuum,
    the expansion  front reaches terminal velocity  immediately and coasts with constant velocity later. The constant terminal velocity in 3-D is the same as in 1-D - the vacuum interface always propagates with the terminal velocity $\gamma = 1+2 \sigma$, independent of the geometry \citep[\cf][]{GreenspanButler}.

    We expect that 
   in the highly relativistic limit,  the structure of the outflow will resemble a relativistic shock wave, where parameters change on a scale $\sim \gamma_{\rm vac}^2$ smaller than the overall size of the out flow.    It is within this narrow region (and even smaller layer near the vacuum interface, see Eqns. (\ref{33}-\ref{341})),  that the self-similar solution derived below is applicable.
   (\cite{GreenspanButler} did consider self-similar 3-D expansion for times much smaller than $r_0/c$, while astrophysical applications require times much larger than $r_0/c$. )
   
   For conical expansion of plasma carrying toroidal \Bf, the conservation laws become
   \ba &&
\partial_t( \gamma \rho) + {1\over r^2} \partial_r ( r^2 \gamma \beta \rho) =0
\nn &&
\partial_t( \gamma B) +{1\over r}  \partial_r (r \gamma \beta B) =0
\nn &&
\partial_t T_{00} + {1\over r^2} \partial_r (r^2 T_{0r}) =0
\nn &&
\partial_tT_{0r}+  {1\over r^2} \partial_r (r^2 T_{rr})=0
%\nn &&
%T_{00}= \gamma^2 ({\cal E}+P + \rho) - P
%\nn &&
%T_{0r}=  \gamma^2 \beta ({\cal E}+P + \rho)
%\nn &&
%T_{rr}=  \gamma^2 \beta^2 ({\cal E}+P + \rho)+P
%\nn &&
%{\cal E}=P = B^2/2
\label{main11}
\ea
where we assumed a constant fixed  polar angle $\theta$ and neglected terms involving $\theta$-dependence.

   In three dimensions,  the continuity and induction equations imply,   $ B= ( \rho r B_0)/(\rho_0 r_0)$ where $r_0$ is an initial radius of the magnetized cavity. 
    This scaling implies that
 for an outflow with a fixed energy,  that starts with a finite region of non-zero \Bf,  
  one cannot assume constant magnetization, since $\sigma \propto  \rho r^2$.   Then, 
the outflow dynamics is not self - similar,  it depends on the initial $
\sigma (r)$. Self-similar solutions  exist  only very close to the edge of the expanding bubble; we derive them next.
   
We assume that at the beginning of the expansion, close to the  leading front, the plasma parameters (density and \Bf) do not vary considerably. Eliminating density in favor of  \Alfven four-velocity $\rho = U_A^2 \rho_0/\sigma (r_0/r)^2$, the condition  that  the magnetic and rest mass energy densities scale similarly with $r$ requires 
   $U_A = g(\chi) /r$. At the vacuum interface the boundary condition is $g=0$.
      Following 
    %Blandford-McKee
     \cite{BlandfordMcKee}, we introduce self-similar variable
   \be
   \chi = (1+2 \gamma_{\rm vac}^2) (1-r/t), \, \chi >1
   \ee and parametrize 
  $\gamma= \gamma_{\rm vac} f$.  (Note that  $ \gamma_{\rm vac}$ is  a constant). 
   In the leading orders in $1/\gamma_{\rm vac}$ we find
   \ba &&
   \chi f g^2 (1+g^2) f' - g (1-\chi f^2 (1+2 g^2)) g'=0
   \nn &&
   (1+ \chi f^2) g f' -2 f (1-\chi f^2) g'=0
   \ea
   The solutions satisfying boundary conditions $f(1)=1$ and $g(1)=0$ are
   \ba &&
   f={1\over \chi^{1/3}}
   \nn &&
   g = {\chi^{1/6}-\chi^{-1/6} 
   \over
     2}
     \ea
     
    % Magnetic field and 
%$B, \, \rho \propto (\chi^{1/6}-\chi^{-1/6})^2 $.

Energy densities per unit range of $\chi$ are given by
\ba &&
U_B d \chi =  (1+\beta^2)/2 \gamma^2 B^2 \propto (1-\chi^{-1/3})^4 d \chi 
\nn &&
U_p  d \chi  = \gamma^2 \rho \propto (\chi^{1/3}-1)^2/\chi  d \chi 
\ea
These relations are applicable in the limit $\chi \rightarrow 1$. To estimate the region of applicability, we equate the total energy in the initial bubble,
$E\sim  B_0^2 r_0^3$ to the energy within the narrow layer near the surface. We integrate $U_B + U_p$ from $\eta =1 $ to $\eta = 1+ \Delta \eta$  to determine $ \Delta \eta$:   $E\sim B_0^2 r_0^2 t (\Delta \eta)^3/\sigma_0^2$. Equating the two,  we find $\Delta \eta$ the width of the self-similar region in coordinate $\eta$
\be
\Delta \eta= \left(  {r_0 \over t} \sigma_0^2 \right)^{1/3}
\label{33}
\ee
In physical coordinates,
\be
\Delta r = \Delta \eta { t \over 2 \Gamma_0^2} = \left(  {r_0  t^2 \over 8 \sigma_0^4 }\right)^{1/3}
\label{341}
\ee
where we used $\Gamma_0 \sim 2 \sigma$. 

  In conclusion of this chapter, we first reiterate that the terminal velocity of expansion into vacuum in 3-D case is the same as in 1D: the terminal Lorentz factor is $\gamma =1 +2 \sigma$ (if starting at rest). The self-similar structure of the resulting bubble is applicable only in a narrow layer near the expanding vacuum interface, in most of the volume the field and velocity structure do depend on the particular initial conditions in the bubble.

\section{Discussion}

In this paper we  found exact  explicit solutions for one-dimensional relativistic   expansion of polytropic fluid into vacuum and into plasma. In particular, 
we discussed an astrophysical important case of 
 strongly magnetized   outflows; in this case especially  simple analytical solutions can be obtained.
   We found exact solutions for one-dimensional  expansion of magnetized plasma into vacuum and into the cold medium both for stationary initial conditions and for a piston moving toward the interface. We found  exact  relations,  applicable for arbitrary magnetization, relativistic motion and  external densities.
These  results can be used for  benchmark estimates of the overall dynamical behavior for the  numerical simulations of relativistic plasmas, \eg\ in heavy ion collisions, and in  strongly magnetized outflows in particular.
 
% Finally, when this paper was close to finishing, a related paper by  \cite{2010arXiv1004.0959G} was submitted. Most of the results  of  \cite{2010arXiv1004.0959G}  for plasma expansion into vacuum   are consistent with our exact solutions. 

 I am greatly  thankful to  Dimitros Gianios, Sergey Komisarov and Alexandre Tchekhovskoy. 

\bibliographystyle{apsrev}
% \bibliographystyle{plain}
% \bibliographystyle{apj}
%\bibliography{~/Home/PulsarRadio/PulsarBib}
%\bibliography{~/Home/Research/BibTex}
%\bibliography{/Users/maxim/Home/Research/BibTex} \end{document}
%\bibliography{~/Home/Research/HallNS}
%\bibliography{/Users/maximlyutikov/Home/Research/HallNS/HallNS}

\end{document}